\documentclass[aps,pra,onecolumn,groupedaddress,showpacs]{revtex4}

\usepackage{hyperref}
\usepackage{graphics}
\usepackage{amssymb,amsmath}
\draft
\usepackage{hyperref}
\usepackage{graphicx}
\newcommand{\Om}{\Omega}

\newcommand{\om}{\omega}
\newcommand{\al}{\alpha}

\newcommand{\ve}{\varepsilon}
\newcommand{\pa}{\partial}

\begin{document}

\title{Optical two-photon nonlinear waves in two-dimensional materials}

\author{G. T. Adamashvili}
\affiliation {Technical University of Georgia, Kostava str.
77, Tbilisi, 0179,  Georgia.\\ email: $guram_{-}adamashvili@ymail.com.$ }

%\date{Received: date / Revised version: date}

\begin{abstract}
A theory of an optical  two-photon breather in a graphene monolayer (or graphene-like two-dimensional material) is constructed.
The system of the material equations for two-photon transitions and the wave equation for transverse magnetic polarized modes of the surface plasmon polaritons are shown to reduce to the nonlinear Schr\"odinger equation with damping. Explicit analytical expressions for a surface small intensity two-photon breather (0$\pi$ pulse) of self-induced transparency are obtained. It is shown that the optical conductivity of graphene reduces the amplitude of the surface two-photon nonlinear wave during the propagation. The one-photon and two-photon breathers in graphene are compared and have obtained that the differences between their parameters are significant.
\end{abstract}


\maketitle

\section{Introduction}

One of the most important consequences of the light-matter interactions are the formation of the nonlinear solitary waves. Depending on the mechanism of the formation of nonlinear waves the resonance and nonresonance waves can be excited. The resonance nonlinear waves  can be investigated within the McCall-Hahn mechanism of the formation of nonlinear waves when a resonance coherent nonlinear  interaction of the wave with optical impurity atoms or semiconductor quantum dots (SQDs) takes place and the conditions of the phenomenon of  self-induced transparency (SIT), $\omega T>>1$ and $ T<<T_{1,2}$ are fulfilled.
Here  $\omega$  and $T$  are the carrier frequency and width of the pulse, respectively. $T_{1}$ and $T_{2}$ are the longitudinal and transverse relaxation times of the resonance optical impurity atoms or SQDs \cite{McCall:PhysRev:69, Maimistov:PhysRep:90, Poluektov:UspFizNauk:74, Allen::75, Adamashvili:PhysRevA:07}. Among SIT nonlinear waves the solitons and breathers are very often met. When a dimensionless of pulse envelope area (energy, in the case of the two-photon SIT) $\Theta$, as a measure of the pulse-atom interaction intensity exceed $\pi$, a soliton is generated, and when $\Theta<<1$, a breather ($0 \pi$ pulse) is excited.
Breather of SIT is of particular interest because  can  be formed for relatively small intensity of the pulse than soliton.  In addition, breather of some physical situations are more stable in comparison to soliton \cite{Chen:Phys. Rev. E :04}. The investigation of stable optical nonlinear waves with a small power in different nanostructures (two-dimensional materials, SQDs, negative index  metamaterials) are among the most promising  research fields in the theory of nonlinear waves (see, for instance, \cite{Adamashvili:PRA:17, Adamashvili:OptLett:06, Adamashvili:Eur.Phys.J.D.:12} and references therein).

The phenomenon of SIT has attracted considerable interest in graphene plasmonics where we can use surface plasmon polaritons (SPPs) \cite{Grigorenko::12}. The properties  of SPPs in the graphene monolayer or graphene-like two-dimensional materials, such as  germanene, silicene, phosphorene, etc. are widely investigated \cite{Geim:Nat.Mater:07, Novoselov::05, Shyam Trivedi::2014, Fan::14, Bao::16, WangLan::16}.
SPP is an  electromagnetic surface wave which can propagate along the flat interface of different materials. The amplitude of the surface wave has a maximum peak at the interface and decays exponentially in both connected samispaces.
SIT for SPPs in various nanostructures  has been investigated and although many impressive results have been obtained, it is still one of most actively researched topics. One major reason is that due to important diverse properties of the resonance nonlinear surface waves in nanostructures, in particular, in graphene monolayer or other two-dimensional materials which are recognized  as being perspective materials for the investigation of nonlinear SPPs and applications \cite{Grigorenko::12}.

The properties of nonlinear SPPs, are especially interesting, when one or several transition layers are sandwiched between the connected materials. The  transition layers can have an influence on the properties of the SPP, especially when they are in resonance with the optical active impurity atoms or SQDs embedded into the transition layer. In such multi-layered system,  resonance solitons and breathers can be created under the condition of SIT in graphene for transverse magnetic (TM) polarized surface and waveguide modes \cite{Adamashvili:PRA:17, Adamashvili:Physica B:14,Adamashvili:Optics and spectroscopy:2015}.

To examine nonlinear SIT waves one has to consider the single-photon and two-photon resonance processes separately and consequently the formations of the single- and two-photon SIT solitons and breathers \cite{Adamashvili:PhysRevE:04,Lopez::06}.
%But in contrast to solitons, the breathers arise in many physical situations when small amplitude optical waves propagate through the medium \cite{Maimistov:PhysRep:90, Poluektov:UspFizNauk:74, Adamashvili:PhysRevE:04, Adamashvili:Technical Physics Letters:14}.
Recently, the surface single-photon SIT breathers in monolayer graphene has been investigated in Ref.\cite{Adamashvili:PRA:17}.
A further study of the properties of the optical nonlinear SPPs in graphene would be the investigation of small intensity resonance two-photon breather in a graphene nanostructure. Such investigation will be able to stimulate the development  a lot of a variety of promising nonlinear two-photon optical processes in two-dimensional materials and we can expect new applications with additional functionalities.

The purpose of the present work is as follows. The investigation of the processes of the formation of optical surface small energy two-photon breather of SIT propagating along the interface between two isotropic media, with two transition layers. One of a transition layers is a graphene monolayer and the second one is described using a model of a two-dimensional gas of  inhomogeneously broadened optical two-photon atoms or SQDs.

\vskip+0.2cm

\section{Basic equations}

We study the excitation of an optical two-photon breather of SIT in the case when an optical the transverse magnetic polarized  pulse with frequency $\omega>>T^{-1}$ and wave vector $\vec{k}=\vec e_{z} k$ is propagating along of the $z$ axis, where  $\vec e_{z}$ is the unit vector along the $z$ axis, $T$ is the width of the surface TM-mode. We investigate the layered structure when a thin transition layer and the graphene monolayer (or graphene-like two-dimensional material) are sandwiched between two isotropic media  with permittivities $\varepsilon_{1}$ and $\varepsilon_{2}$, respectively. A thin transition layer of thickness $h<<\lambda$ which contains a small concentration of noninteracting SQDs or optical impurity atoms, of density $n_{0}$, where $\lambda$ is the wavelength of the surface TM-mode. We approximate  the graphene monolayer and the transition resonance layer, each to be infinitely
thin, in which case we can approximate both by the Dirac delta function $\delta(x)$. The two-photon polarization of the resonance transition layer $\vec{P}(x=0,z,t)=\vec e_{z}\;p_{2}(z,t)$, is determined by the ensemble of optical active impurity two-photon atoms (or SQDs). The electric current density of the graphene monolayer (at $x=0$) is given by $\sigma \; \vec{E}(z,t)$, where $\sigma$ is the electrical conductivity of graphene, $ \vec{E}$ is the vector of the strength of the electrical field for a SPP.

In order to consider two-photon excited processes we present  a Fourier-decomposition of the $z$-components of the
vector of the strength of the electrical field $\vec{E}(E_{x},0,E_{z})$  of the surface TM-mode in the connected media in the following form:
\begin{equation}\label{Fu1}
{E}_{{z;}_{j}}(x,z,t)=\int \int \mathcal{E}_{{z;}_{j}}(\tilde{\Omega},\tilde{Q})e^{\vartheta_{j} \kappa_{j}(\tilde{\Omega},\tilde{Q}) x}e^{i(\tilde{Q}z-\tilde{\Omega} t)}
d\tilde{\Omega} d\tilde{Q},
\end{equation}
where
\begin{equation}\label{kappa}
\kappa_{j}^{2}(\tilde{\Omega}, \tilde{Q})=\tilde{Q}^{2}-  \varepsilon_{j} \frac{\tilde{\Omega}^{2}}{c^{2}},
\end{equation}
$j=1,2;\;\;$ $\mathcal{E}_{{z;}_{1}}(\tilde{\Omega},\tilde{Q})=\mathcal{E}_{{z;}_{2}}(\tilde{\Omega},\tilde{Q}),\;\;
$$\vartheta_{1}=1,\;\vartheta_{2}=-1;\;\;$  $x<0$: medium 1 and $x>0$: medium 2, $c$ is the velocity of light in vacuum.
Eqs. \eqref{Fu1} and \eqref{kappa} are determine the transverse structure of the surface TM mode.

The nonlinear wave equation for the z-component of the strength of the electrical field for a SPP at  $x=0$ has the form \cite{Adamashvili:Physica B:14,Adamashvili:PRA:17}
\begin{equation}\label{eq1}
(A+ iB \frac{\pa }{\pa t}-iC \frac{\pa }{\pa z}-a \frac{\pa^2 }{\pa t^2}
+d\frac{\pa^2 }{\pa t \pa z} -b \frac{\pa^2 }{\pa z^2}) E_{z}=-4\pi  p_{2}-{4\pi \sigma}
\int E_{z} dt
\end{equation}
where we use the notations
\begin{equation}\nonumber\\
A=f(\om,k)-\om f_{\Om} -k f_{Q} +a \om^{2} + d k \om  + b k^{2},
$$$$
B=f_{\Om} - 2 a \om  - d k,\;\;\;\;\;\;\;\;\;\;\;
C=f_{Q} - d \om  - 2 b k,
\end{equation}
$$
f_{\Om}=\frac{\pa f}{\pa
\tilde{\Om}}|_{\tilde{\Omega}=\om,\tilde{Q}=k},\;\;\;\;\;\;\;\;\;\;\;f_{Q}=\frac{\pa f}{\pa
\tilde{Q}}|_{\tilde{\Omega}=\om,\tilde{Q}=k},\;\;\;\;\;\;\;\;\;\;\;\;\;
a=\frac{1}{2} \frac{\pa^{2} f}{\pa \tilde{\Omega}^{2}}|_{\tilde{\Omega}=\om,\tilde{Q}=k},$$$$
b=\frac{1}{2} \frac{\pa^{2} f}{\pa \tilde{Q}^{2}}|_{\tilde{\Omega}=\om,\tilde{Q}=k},\;\;\;\;\;\;\;\;\;\;\;
d=\frac{\pa^{2} f}{\pa \tilde{Q} \pa \tilde{\Omega}}|_{\tilde{\Omega}=\om,\tilde{Q}=k},\;\;\;\;\;\;\;\;\;\;\;
f(\tilde{\Omega},\tilde{Q})= \frac{\ve_{1}}{\kappa_{1}}+\frac{\ve_{2} }{ \kappa_{2}}.
$$
In Eq.\eqref{eq1} the last two terms are the contributions from the resonance two-photon transition layer and the graphene monolayer.

We can simplify Eq.\eqref{eq1}  using the method of slowly varying  envelope \cite{McCall:PhysRev:69, Maimistov:PhysRep:90, Poluektov:UspFizNauk:74, Allen::75}. For this, we represent the $z$-component of the electric field  of the surface wave in the form
\begin{equation}\label{slow}
 E_{z}=\sum_{l=\pm 1}\hat{E}_{l}Z_{l},\;\;\;\;\;\;\;\;\;\;\;\;Z_{l}= e^{il(kz -\om t)},
\end{equation}
where $\hat{E}_{l}$ is the slowly changing complex profile of the electric field of a SPP.
We suppose that the inequalities
\begin{equation}\label{sl}\nonumber\\
|\frac{\partial \hat{E_{l}}}{\partial t}|<<\omega
|\hat{E_{l}}|,\;\;\;|\frac{\partial \hat{E_{l}}}{\partial z }|<<k|\hat{E_{l}}|
\end{equation}
are fullfilled.

Substituting Eq.\eqref{slow}  into the wave equation \eqref{eq1}, we obtain
\begin{equation}\label{sl2}
 Z_{+1} \{ i f_{\Om} \frac{\pa }{\pa t}
  -i f_{Q} \frac{\pa }{\pa z}
  +d \frac{\pa^{2} }{{\pa t}{\pa z}}
 - a \frac{\pa^{2} }{\pa t^{2}}
-b\frac{\pa^{2} }{\pa z^2} \}\hat{E}_{+1} +\mathcal{O}(Z_{-1})
=-4\pi  p(z,t)-{4\pi \sigma}
\int E_{z} dt
\end{equation}
and dispersion equation in the form
\begin{equation}\label{disp}
k^{2}=\frac {\om^{2}}{ c^{2}}\;
\frac{\varepsilon_{1}\varepsilon_{2}}{\varepsilon_{1}+\ve_{2}}.
\end{equation}

For the further transformation of Eq.\eqref{sl2} we can use of the perturbative reduction method \cite{Taniuti::1973}, according to which the function  $\Psi_{l}(z,t)=\int_{-\infty}^{t}\hat{E}_{l}(z,t') dt'$ can be represented in the form
\begin{equation}\label{psi}
\Psi_l(z,t)=\sum_{\alpha=1}^{\infty}\varepsilon^\al \Psi^{(\alpha)}_{l}=\sum_{\alpha=1}^{\infty}\sum_{n=-\infty}^{+\infty}\varepsilon^\al
Y_n f_{l,n}^ {(\alpha)}(\zeta,\tau),
\end{equation}
where
$$
Y_n=e^{in(Qz-\Omega t)},\;\;\;\zeta=\varepsilon Q(z-v_g
t),\;\;\;\tau=\varepsilon^2 t,\;\;\;\  v_g=\frac{d\Omega}{dQ}.
$$

This method allow us to expand the function $\Psi_l$ in the more slowly changing functions $ f_{l,n}^{(\alpha )}$. Therefore supposed  to take place the inequalities:
\begin{equation}\label{rtyp}\nonumber\\
\omega\gg \Omega,\;\;k\gg Q,\;\;\;
\end{equation}
$$
\left|\frac{\partial
f_{l,n}^{(\alpha )}}{
\partial t}\right|\ll \Omega \left|f_{l,n}^{(\alpha )}\right|,\;\;\left|\frac{\partial
f_{l,n}^{(\alpha )}}{\partial z }\right|\ll Q\left|f_{l,n}^{(\alpha )}\right|.
$$

The dependence of the two-photon polarization $p_2=n_{0} E_{z} \sum_{l=\pm 1}B_{l}Z_{2l}$  on the strength of
the electrical field is governed by the optical two-photon material equations
\begin{equation}\label{tbl}
\frac{\partial   B_{\pm 1}  }{\partial t} =\pm i (\Delta  +\frac{r_{22}-r_{11}}{4\hbar}\hat{E}_{+1}\hat{E}_{-1} ) B_{\pm 1}  \pm i\frac{{\kappa }_{0}}{2} \hat{E}_{\pm 1}^{2} N,
$$$$
\frac{\partial N}{\partial t}=-i{\kappa }_{0} (B_{-1} \hat{E}_{+1}^{2} - B_{+1}\hat{E}_{-1}^{2}),
\end{equation}
where
$$
{\kappa }_{0}=\frac{|r_{21}|^{2}}{2\hbar},\;\;\;\;\;\;\;\;\;\;\;\;\;\;\;\;\;\;\;\;\;\;\;\;
\Delta=2\omega-\omega_{0}
$$
$$
r_{21}=r_{12}^{*}=\sum_{m}\frac{\mu_{1m}\mu_{m2}}{\hbar(\om_{m2}
 +\om)},\;\;\;\;\;\;\;\;\;\;\;\;\;\;\;\;\;\;\;\;\;\;\;\;
 r_{ii}=\frac{2}{\hbar}\sum_{m }\frac{|\mu_{im}|^{2}\om_{mi}}
 {\om_{mi}^{2}-\om^{2}},\;\;\;\;\;\;\;\; (i=1,2),
$$
$4B_{+1}B_{-1}+ N^{2}= 1,\;\;$$\hbar$ is the Planck's constant, $\om_{nm}$ and $\mu_{nm}$ are the frequencies and matrix elements of electric-dipole moments transitions between n and m level of energy of the impurity optical atoms or SQDs \cite{Maimistov:PhysRep:90, Poluektov:UspFizNauk:74, Adamashvili:PhysRevE:04, Adamashvili:Technical Physics Letters:14}.

Substituting Eq.\eqref{slow} into the optical two-photon material equations  \eqref{tbl}, and taking into account \eqref{psi},
under the condition of the inhomoheneouse broadening of the spectral line, we obtain the two-photon polarization in the following form \cite{Bloembergen::79}
\begin{equation}\label{pol}
p_{2}=  i \ve^{3} \frac{{\kappa }_{0}}{2}n_{0}\int \frac{g(\Delta)d\Delta}{1+T^2 \Delta^2} \sum_{l=\pm 1}l  Z_{l} \frac{\pa \Psi^{(1)}_{-l}}{\pa t} \int_{-\infty}^{t} (\frac{\pa \Psi^{(1)}_{l}}{\pa t})^{2}  dt',
\end{equation}
where $g(\Delta)$ is the inhomogeneous broadening lineshape function for an ensemble of two-level optical atoms or SQDs.

Substituting Eqs.\eqref{psi} and \eqref{pol} into the wave equation \eqref{sl2}, we obtain the nonlinear wave equation in the following form

\begin{equation}\label{hhd}
 \sum_{\alpha=1}^{\infty}\sum_{n=-\infty}^{+\infty}\varepsilon^\al  Z_{+1}
Y_{+1,n}  \{
{w}_{+1,n}
+\varepsilon j_{+1,n}
\frac{\partial }{\partial \zeta}
+\varepsilon^2 h_{+1,n}\frac{\partial }{\partial \tau}
+i\varepsilon^{2}H_{+1,n} \frac{\partial^{2} }{\partial \zeta^{2}}\}f_{+1,n}^{(\alpha)}
=$$$$=-i \ve^{3} R  Z_{+1} [ Y_{+1}  |f_{+1,+1}^ {(1)}|^{2} f_{+1,+1}^ {(1)}
+  Y_{-1}  |f_{+1,-1}^ {(1)}|^{2} f_{+1,-1}^ {(1)}]$$$$- \varepsilon \tilde{\sigma} Z_{+1}
(Y_{+1} f_{+1,+1}^ {(1)}+Y_{-1} f_{+1,-1}^ {(1)})
+O(Z_{-1}),
\end{equation}
where
\begin{equation}\label{hh}
{w}_{+1,n}=
-in\Omega (n f_{\Om} {\Omega}+n f_{Q}  Q  +  {\Omega} Q d + a   {\Omega}^{2}+b  Q^{2}  ),
$$$$
j_{+1,n}=-Q[ 2 f_{\Om} n \Omega  v_g  + f_{Q} n(Q v_g +\Omega)+  d   \Omega ({\Omega} +  2   Q v_g )+ 3a {\Omega}^{2}   v_g + b Q(Q v_g +  2 \Omega)],
$$$$
h_{+1,n}=2 n f_{\Om} \Omega + n f_{Q} Q  + 2 d Q \Omega + 3 a {\Omega}^{2}+ b Q^{2},
$$$$
H_{+1,n}= Q^{2}[ f_{\Om}  v_g^{2}+  f_{Q}  v_g + n d     v_g ( 2  \Omega
+   Q v_g )+ 3 n a  \Omega  {v_g}^{2}  + n b ( 2Q v_g +\Omega)],
$$
$$
R=\frac{\pi n_{0} |r_{21}|^{2}\Omega^{2}}{2 \hbar} \int \frac{g(\Delta)d\Delta}{1+T^2 \Delta^2},
$$
$$
\tilde{\sigma}=\frac{{4\pi \sigma}\Omega}{\om}=\varepsilon^{2}\Gamma.
\end{equation}

\vskip+0.2cm

\section{Two-photon breather solution}

To find the values of $f_{+1,n}^{(\alpha)}$ from Eq. \eqref{hhd} we equate to each other the terms corresponding to the same powers of $\varepsilon$. This leads to a set of equations. As a result, we obtain that $j_{+1,n}=0$ and only  components of $f _{+ 1,+ 1}^{(1)}$ and $f _{+ 1,-1}^{(1)}$  are differ from zero.

The relation between the parameters $\Omega $ and $Q$ is determined from Eq.\eqref{hh} and has the form
\begin{equation}\label{hj}
n f_{\Om} {\Omega}+ n f_{Q} Q   + a   {\Omega}^{2}+b Q^{2} +d  {\Omega}Q =0
\end{equation}
and also we get the equation
\begin{equation}\label{rty3}
v_{g}=- \frac{n f_{Q} +  d   \Omega +  2b  Q }{  n f_{\Om} + d  Q  + 2a {\Omega}}.
\end{equation}
From Eq.\eqref{hhd}, to third order in $\varepsilon $, we obtain the following nonlinear equation
\begin{equation}\label{eq7}
i \frac{\partial f_{+1,\pm 1}^ {(1)}}{\partial \tau}-
 \frac{H_{+1,\pm 1}}{h_{+1,\pm 1}} \frac{\partial^{2} f_{+1,\pm 1}^ {(1)}}{\partial
\zeta^{2}}-  \frac{R}{h_{+1,\pm 1}}  |f_{+1,\pm 1}^ {(1)}|^{2} f_{+1,\pm 1}^ {(1)}+ i  \frac{\Gamma}{h_{+1,\pm 1}} f_{+1,\pm 1}^ {(1)}=0.
\end{equation}

Eq.\eqref{eq7}  will be transformed to a damped nonlinear Schr\"odinger equation  in the form
\begin{equation}\label{eq9}
i\frac{\partial \Lambda}{\partial t}-\frac{\partial^{2}
\Lambda }{\partial y^2}- |\Lambda|^{2}\Lambda=-i \gamma \Lambda,
\end{equation}
where
$\Lambda =\varepsilon \sqrt{q}
f_{+1,+1}^{(1)},\;\;$$i \gamma \Lambda$ is the damping term,$\;\;\;\;
y=\frac{1}{\sqrt{p}}(z-v_g t),\;\;\;t=t,$
$$
p=\frac{H_{+1,+1}}{h_{+1,+1} Q^2},\;\;\;\;\;\;\;\;\;\;\;\;\;\;
q= \frac{R}{h_{+1,+1}},\;\;\;\;\;\;\;\;\;\;\;\;\;\;
\gamma=\frac{\tilde{\sigma}}{h_{+1,+1}}.
$$

The solution of Eq.\eqref{eq9} has the following form \cite{Adamashvili:Phys.Rev.A:14, Ablowitz::81, Adamashvili:PRA:17, Adamashvili:Phys.Rev.A:08}:
\begin{equation}\label{eq12}
\Psi_{+1}(z,t)=- \frac{4\eta}{\sqrt{q}}\frac{\sin (Qz-\Omega t -\varphi _{1})}{\cosh2\eta
\varphi _{2}}  +{\cal O}(\varepsilon^2) .
\end{equation}
where
$$
\varphi _{1}=\frac{2\xi z}{\sqrt{p}}+2[2({\xi
}^{2}-\eta ^{2})-\frac{{\xi
}v_{g}}{\sqrt{p}}]t-\varphi _{0},
$$
\begin{equation}\nonumber\\
\varphi _{2}=\frac{z}{\sqrt{p}}+(4\xi -\frac{v_{g}}{\sqrt{p}
})t-y_{0}.
\end{equation}
The parameters $\xi ,\eta ,\varphi _{0}$ and $y_{0}$ are the
scattering data. These quantities arise when the nonlinear equation is analyzed by the inverse
scattering transform \cite{Ablowitz::81}.

The equation \eqref{eq12} describes the two-photon breather for surface TM-mode at $x=0$.
The evolution of the two-photon breather amplitude $\frac{4 \eta } {\sqrt{q}}$  due to the influence of the graphene conductivity is determined from the equation
\begin{equation}\label{eta}
\eta (t)=\eta(0)e^{-2 t \gamma},
\end{equation}
where $\eta(0)$ is the initial value of $\eta$ at t=0.

\vskip+0.2cm
\section{Conclusion}

We study the processes of formation surface TM-modes at the interface of the two isotropic media with an ensemble of SQDs (impurity optical atoms) transition layer and graphene monolayer sandwiched between the two connected media. We have shown that in the propagation of optical pulse through such layered system under the condition of SIT  the optical two-photon breather of SPP can arise. The explicit form and parameters of the optical surface two-photon breather for any value of $x,z$, and $t$  are determined from Eqs.\eqref{Fu1}, \eqref{kappa}, \eqref{hh}, \eqref{rty3}, \eqref{eq12} and \eqref{eta}. The dispersion equation and the relation between quantities $\Omega$ and $Q$ are given by Eqs.\eqref{disp} and \eqref{hj}, respectively. From these equations we can see that the parameters of the surface optical two-photon breathers depend on the parameters of the optical atoms (or SQDs) $R$, well as on the connected media $\ve_{1,2}$ and also depend on the transverse structure of the surface TM-mode through the Eqs.\eqref{Fu1} and \eqref{kappa}. The amplitude of the two-photon breather can be expected to decay exponentially, in the process of propagation [Eq.\eqref{eta}].

Having presented the theoretical results for the  two-photon SIT breather in graphene, we will be able to compare some properties of two-photon breather with one-photon breather investigated earlier in Ref.\cite{Adamashvili:PRA:17}.

For theoretical investigations of SIT for the wave equation it is sufficient to take into account only  the first derivatives of $\hat{E}$ with respect to the space coordinates and time. The corresponding second derivatives usually  are ignored. Such situation take place as for one- and two-photon solitons and so one-photon breathers as well \cite{McCall:PhysRev:69, Maimistov:PhysRep:90, Poluektov:UspFizNauk:74, Allen::75, Adamashvili:PhysRevA:07, Adamashvili:PRA:17}.
But for two-photon breathers situation is different.
Indeed, the polarization of the optical atoms (or SQDs) under the condition of the two-photon processes $p_{2}$ is  of order $\ve^3$, [Eq.\eqref{pol}].  Consequently, unlike one-photon polarization, which  has as linear and nonlinear parts, the two-photon polarization has only nonlinear part. This is a very significant difference. In particular, unlike one-photon breather \cite{Adamashvili:PRA:17}, for two-photon breather the relation between slow oscillating parameters $\Omega$ and $Q$ [Eq.\eqref{hj}] and the quantity $v_{g}$  [Eq.\eqref{rty3}] do not depend on the coefficient of resonance optical absorption $R$ [Eq.\eqref{hh}]. Therefore, if we neglected the second derivatives in the wave equation Eq.\eqref{eq1}, i.e. if we take $a=b=d=0,\;$ $v_{g}=-f_{Q}/f_{\Omega},\;$ then we have $H_{+1,+1}=0$ and under this condition nonlinear Schr\"odinger equation  [Eq.\eqref{eq7}] has not the two-photon breather solution. This circumstance leads to the fact that in contrast of the one-photon processes,  the two-photon breather solution we can obtain only under the condition if we to  take into account besides of the first derivatives of the function $E$ also the second derivatives in Eqs.\eqref{eq1},\eqref{sl2},\eqref{hhd} and \eqref{hj}. As a result, all characteristic equations and parameters for one- and two-photon breathers are absolutely different and consequently, their properties will be different too (see, for comparison Ref.\cite{Adamashvili:PRA:17}).

Having presented the theoretical results for graphene, after corresponding transformations, we can use these results also for other two-dimensional materials \cite{ Shyam Trivedi::2014, Fan::14, Bao::16, WangLan::16}. Because the phenomenon of two-photon excitation finds applications in various technical areas,  we can expect that many new applications for two-photon SIT breathers will also take place in two-dimensional materials.

\vskip+0.02cm
\section{acknowledgments}

This research was supported by the Georgian Shota Rustavely NSF Grant No. 217064.

\end{document}